\documentclass[reprint,showkeys,aip,jap]{revtex4-1}

\usepackage[pdftex]{graphicx}% Include figure files
\usepackage{dcolumn}% Align table columns on decimal point
\usepackage{bm}% bold math

\begin{document}

\title{Collective behavior of bulk nanobubbles produced by the alternating polarity electrolysis}

\author{Alexander V. Postnikov}
\affiliation{Yaroslavl Branch of the Institute of Physics and Technology RAS, 150007, Universitetskaya 21, Yaroslavl, Russia}

\author{Ilia V. Uvarov}
\affiliation{Yaroslavl Branch of the Institute of Physics and Technology RAS, 150007, Universitetskaya 21, Yaroslavl, Russia}

\author{Nikita V. Penkov}
\affiliation{Institute of Cell Biophysics, RAS, 142290 Pushchino, Moscow Region, Russia}

\author{Vitaly B. Svetovoy}
\email[Corresponding author: ]{v.b.svetovoy@rug.nl}
\affiliation{Yaroslavl Branch of the Institute of Physics and Technology RAS, 150007, Universitetskaya 21, Yaroslavl, Russia}
\affiliation{Zernike Institute for Advanced Materials, University of Groningen - Nijenborgh 4, 9747 AG Groningen, The Netherlands}

\begin{abstract}
Nanobubbles in liquids are mysterious gaseous objects having exceptional stability. They promise a wide range of applications but their production is not well controlled and localized. Alternating polarity electrolysis of water is a tool that can control production of bulk nanobubbles in space and time without generating larger bubbles. Using the schlieren technique a detailed three-dimensional structure of a dense cloud of nanobubbles above the electrodes is visualized. It is demonstrated that the thermal effects produce different schlieren pattern and have different dynamics. A localized volume enriched with nanobubbles can be separated from the parent cloud and exists on its own. This volume demonstrates buoyancy from which the concentration of nanobubbles is estimated as $2\times 10^{18}\:$m$^{-3}$. This concentration is smaller than that in the parent cloud. Dynamic light scattering shows that the average size of nanobubbles during the process is $60-80\:$nm. The bubbles are observed 15 minutes after switching off the electrical pulses but their size is shifted to larger values of about $250\:$nm. Thus, an efficient way to generate and control nanobubbles is proposed.
\end{abstract}

\maketitle

%%%MAIN TEXT%%%%
\section{Introduction}
In spite of a long history of water electrolysis \cite{deLevie1999} and its applications \cite{Zeng2010} the process performed by short voltage pulses of alternating polarity brought many surprises \cite{Svetovoy2016}. The Faraday current density was as high as $100\:$A/cm$^2\ $ \cite{Svetovoy2013} while the highest value observed in ordinary DC electrolysis is $1\:$A/cm$^2\ $ \cite{Vogt2005}. Despite the high current density no visible bubbles are produced in the process \cite{Svetovoy2011} although high concentration of gases in the electrolyte has been proven by different methods: a significant reduction of the refractive index of liquid \cite{Svetovoy2011} and a considerable pressure increase in a closed microchamber \cite{Svetovoy2014} were observed. The visible bubbles reappear in the system when a stoichiometric ratio $2:1$ between hydrogen and oxygen is broken locally, for example, if one electrode is kept different times at positive and at negative potential.

Large current and high frequency ($\sim 100\:$kHz) of the polarity change results in an extreme supersaturation ($\sim 1000$) with both gases in the vicinity of the electrodes \cite{Svetovoy2013}. This is much higher than the maximal value $<100$ observed in the DC electrolysis \cite{Vogt1993}. The supersaturation is high enough to overcome the barrier for homogeneous nucleation of the bubbles  \cite{Debenedetti1996,Skripov1974} that was observed at special conditions as a faint haze above the entire area of the electrodes \cite{Svetovoy2013}. When the positive and negative pulses have exactly the same duration the bubbles do not grow large and stay below $200\:$nm.

Direct observation of NBs is complicated by a small size and short lifetime of the objects. In the closed chamber with dimensions of $100\times 100\times 5\:\mu$m$^3$ the pressure is released in $100\:\mu$s \cite{Svetovoy2014} that can be considered as a typical lifetime of the bubbles. The short lifetime seems a distinctive feature of NBs produced by the alternating polarity electrochemical process, but here we will see that the bubbles can live much longer.

On the other hand, NBs produced mechanically are actively discussed in the last decade \cite{Alheshibri2016,Lohse2015}. These nanoscopic gaseous domains attracted significant attention because they live unexpectedly long. The theory of diffusive dissolution predicts the lifetime in the microsecond range \cite{Epstein1950} while the NBs live hours or even days. The reason for this exceptional stability is still under discussion. One must distinguish between surface NBs \cite{Lohse2015}, which exist at the liquid-solid interface, and bulk NBs \cite{Alheshibri2016} existing in the liquid volume. The latter are investigated in less degree but they are relevant to this study.

Ohgaki {\it et al.} \cite{Ohgaki2010} produced the bulk NBs with a rotary pump. Supersaturation with nitrogen as high as 36 was reached and water density was reduced in accordance with the gas content. The bubbles were visualized using freeze-fracture replicas observed with a scanning electron microscope. The bubbles with an average size of $100\:$nm lived more than two weeks. Using a similar mechanical method to produce NBs Ushikubo {\it et al.} \cite{Ushikubo2010} measured the size distribution of oxygen NBs with the dynamic light scattering (DLS). Furthermore, they measured the $\zeta$-potential of the bubbles in a range of $20-40\:$mV and related the stability of NBs with the charges on their walls. An advanced theoretical model of the bubble stability due to charging was developed by Yurchenko {\it et al.} \cite{Yurchenko2016}. Using different optical techniques Bunkin {\it et al.} \cite{Bunkin2012} determined sizes of NBs appearing in sodium chloride solution without external stimuli. They have been able to distinguish latex nanoparticles from NBs optically.

Bulk NBs produced by the DC electrolysis of water were described in a series of papers by Kikuchi {\it et al.} \cite{Kikuchi2001,Kikuchi2006a,Kikuchi2006b,Kikuchi2007,Kikuchi2009}. Hydrogen content of alkaline water was measured using a dissolved hydrogen meter and using a chemical method \cite{Kikuchi2001}. Difference in the results was explained by the existence of NBs, which do not contribute to the dissolved hydrogen concentration. Independent DLS measurement supported presence of NBs. Similar conclusions were made analyzing water from the anode chamber, which is supersaturated with oxygen \cite{Kikuchi2009}. It was found that hydrogen NBs live a few hours while oxygen NBs live a few days.

A wide range of applications is expected for NBs. They have been applied for nanoscopic cleaning \cite{Wu2008,Liu2008,Liu2009,Zhu2016}, for waste water treatment \cite{Tasaki2009,Agarwal2011}, for medical \cite{Mondal2012,Modi2014,Owen2016} applications, and for heterocoagulation \cite{Mishchuk2006,Azevedo2016}.

When the NBs are produced by the alternating polarity electrolysis the source of the bubbles is localized at the electrodes. Existence of a cloud of NBs above the electrodes was demonstrated in a recent paper \cite{Postnikov2017}. A high concentration of gases was estimated from the optical distortion of the electrodes but only indirect evidence that the gas is contained in the NBs was presented. In this paper using the schlieren technique we visualize the space distribution and dynamic behavior of NBs. We demonstrate that NBs can form lumps living separately from the source. Bubble size and their lifetime are measured using the DLS method.

\section{Experimental}

\subsection{Samples and process}
\begin{figure}[tbhp]
\centering
\includegraphics[width=.95\linewidth]{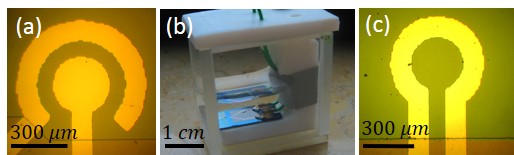}%Samples
\caption{(a) Titanium electrodes, the contact lines are insulated with SU8. (b) Cuvette with a sample inside covered by a layer of the electrolyte. (c) The heating element. The chromium loop is open but the contact lines are covered with SU8. }
\label{fig:electrodes}
\end{figure}

Electrodes are fabricated on an oxidized silicon wafer. All metallic layers are deposited by magnetron sputtering. First, $10\:$nm Ti adhesion layer is deposited. Next, $500\:$nm Al layer is deposited to minimize the resistance of contact lines. Finally, $100\:$nm thick working layer of titanium is sputtered. The electrodes and contact lines are patterned with the lift-off process. The contact lines are insulated with $8\:\mu$m thick SU8 photoresist. The wafer is diced on separate samples with sizes of $2\times 1\:$cm$^2$ and $1\times 1\:$cm$^2$. Concentric shape of the electrodes (see Fig.$\:$\ref{fig:electrodes}(a)) is chosen for better concentration and symmetric distribution of NBs. Titanium as a material for electrodes demonstrates stability at high current density. A disadvantage of Ti electrodes is tendency to oxidation that increases voltage applied to the electrodes.

A sample is fixed at the bottom of a cuvette (see Fig.$\:$\ref{fig:electrodes}(b)) with dimensions of $3\times 2\times 3\:$cm$^3$ ($l\times w\times h$). Solution of Na$_2$SO$_4$ in distilled water at a concentration of $1\:$mol/l is used as an electrolyte. The solution layer above the electrodes can be in the range $2-10\:$mm. Square voltage pulses of alternating polarity with the amplitude $U$ and frequency $f$ (typically $f=200\:$kHz) are applied from the internal generator of a PicoScope 5000 and amplified by a homemade amplifier. A typical voltage used in this experiment is $11\:$V. The pulses can be generated continuously with a constant amplitude or with the amplitude modulated by a triangle profile. The voltage, current through the cell, and triggering pulses for camera are recorded by the PicoScope.

\subsection{Schlieren setup}
\begin{figure}[tbhp]
\centering
\includegraphics[width=.95\linewidth]{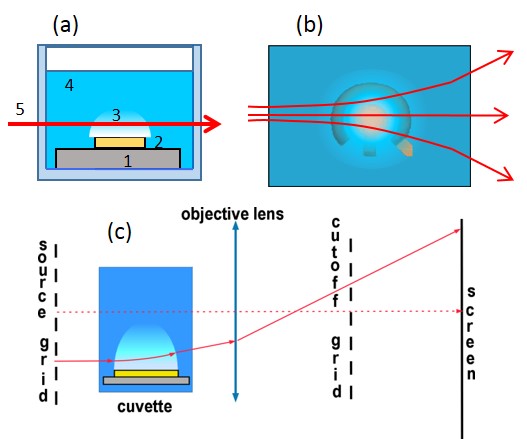}%Scheme
\caption{Scheme of the schlieren experiment. (a) Sample in the cuvette (side view). Substrate, electrodes, cloud of NBs, electrolyte, and light beam passing through the cloud are numbered from 1 to 5, respectively. (b) Top view on the electrodes covered with the cloud of NBs. Refraction of the light beam is shown. (c) Scheme of the lens and grid schlieren setup.}
\label{fig:scheme}
\end{figure}

The lens and grid schlieren  technique \cite{Weinstein2010} is used in this work to visualize variation of the optical density  induced by the electrochemical process. The system uses a one dimensional array (grid) of light sources instead of a point source in the conventional schlieren system. The source grid is imaged with an objective onto a one dimensional array of corresponding knife-edge cutoffs. The grid is a series of alternating opaque and transparent stripes. The cutoff grid is a photographic negative image of the source grid. The system is shown schematically in Fig.$\:$\ref{fig:scheme}. It can be thought of as a superposition of many conventional schlieren systems, with each transparent stripe of the source grid and corresponding opaque stripe of the cutoff grid functioning as the light source and knife edge, respectively \cite{Florian2012}.

To increase the sensitivity the transparent stripes in the source grid have to be narrow but the diffraction effects may become important blurring the image \cite{Weinstein2010}. The field of view is limited by the source grid and viewing angle. If the numerical aperture of the imaging lens is large enough the system will have a narrow depth of focus and such a system is described as focusing schlieren  system. Orientation of the grids (horizontal or vertical) depends on the desired direction of sensitivity.

The source grid of our schlieren setup consist of a series of $400\:\mu$m opaque stripes separated by $100\:\mu$m transparent gaps. For the cutoff grid the stripes are inverted. The grids are $40\:$mm in diameter quartz substrates sputtered with a $100\:$nm thick chromium layer. The substrate is patterned with standard lithography. The cuvette with electrolyte is positioned in between the source grid and the lens.  A lens with $50\:$mm focal length and with $25\:$mm aperture is used to image inhomogeneities of the optical density. The grids are aligned with a horizontal microscope. If the system is properly adjusted the sensor is illuminated uniformly. The cutoff grid is aligned so that the regions of lower refraction becomes darker and the regions of higher refraction becomes brighter compared to background.

\subsection{Thermal effect}
Temperature of the electrolyte during the process is measured with a copper-constantan thermocouple. For insulation the wires and the junction ($200\:\mu$m in size) are covered with a compound OMNIVISC 1050 (thickness $10-20\:\mu$m).

A special heater (see Fig.$\:$\ref{fig:electrodes}(c)) has been fabricated on silicon substrate to mimic the Joule heating produced by the current in the electrochemical process. The heater is a $100\:$nm thick chromium loop that has the same shape as the gap between the electrodes. The resistance of the loop is $3.63\:\Omega$ at $27^{\circ}\:$C. The contact lines are covered by $8\:\mu$m of SU8 photoresist and the total resistance of the heating element is $14\:\Omega$. The thermal coefficient $\alpha=2.4\times 10^{-3}\:$C$^{-1}$ of the heater was found by measuring the resistance in distilled hot water. The heater has four contact lines that allows the voltage application and the measurement of the voltage drop on the loop independently.

\subsection{Dynamic light scattering}
The size distribution of the NBs is determined using the dynamic light scattering method. The measurement is made with Zetasizer nano ZS analyzer (Malvern, UK). The samples are placed in a polystyrene cuvette with a cross-section of $10\times 10\:$mm$^2$. The cuvette is filled with the electrolyte. A laser beam (wavelength $633\:$nm) has a gaussian width of $0.1\:$mm. The scattering angle is fixed and equal to $173^{\circ}$ (backscattering).

The electrolyte was thoroughly filtered by a filter with an average pore diameter of $200\:$nm before the use. We did measurements only if the solution did not demonstrate any particles before the bubble generation. The bubbles are generated at a distance of $5-7.5\:$mm from the bottom of the cuvette and the laser beam passes $8\:$mm above the bottom. Due to the instrument design it is difficult to control precisely position of the beam above the electrodes. For this reason the beam could be not in the optimal position (above the central electrode) but blind shifts of the cuvette within $2\:$mm do not change the signal significantly.

The measurements are done as follows. First, the electrolyte before the bubble generation is measured to check absence of external nanoparticles in the solution. Then the bubbles are generated during $2\:$min by applying continuous voltage pulses with an amplitude of $8-12\:$V at frequencies 150 and $325\:$kHz. One measurement is made during the generation. When the electrical pulses are switched off a few more measurements are made during $15\:$min after the generation.

\section{Results and discussion}

\subsection{Faraday current}
\begin{figure}[tbhp]
\centering
\includegraphics[width=.95\linewidth]{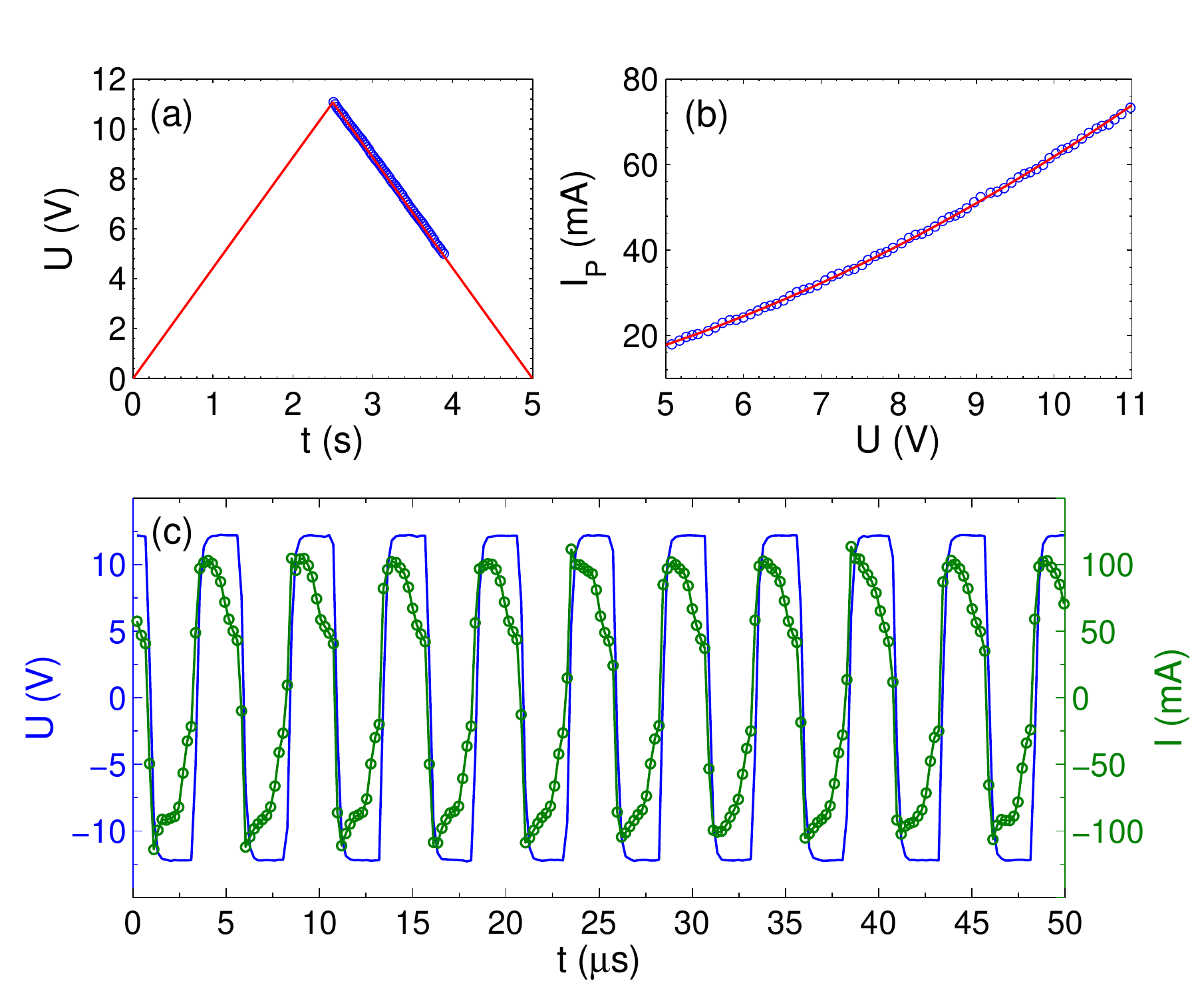}%Current_50_H.pdf
\caption{Voltage and current through the cell. (a) Amplitude of applied voltage pulses as a function of time. Blue circles mark the time interval, where high-resolution data were collected. (b) The current is determined from the power (see text) as a function of applied voltage amplitude $U$. (c) Time-resolved voltage (left axis) and current (right axis) pulses. The open circles show the actual sampling.}
\label{fig:current}
\end{figure}

If a DC voltage or single polarity pulses are applied to the electrolytic cell, one electrode is oxidized fast and the process stops leaving a few visible gas bubbles. When alternating polarity pulses are applied a significant current is flowing through the electrolyte as shown in Fig.$\:$\ref{fig:current} but no visible gas is observed. In this case an oxide layer is formed on one electrode during half of the period but then during the second half it is reduced. In contrast with such metals as Pt, Cu, or W the current contains not only the charging-discharging and the Faraday components \cite{Svetovoy2013}, but also includes the oxidation-reduction component. For this reason the Faraday current $I_F$ cannot be extracted from each pulse using a simple 3-parametric fit $I(t)=I_F+I_1e^{-t/\tau}$ as it was done in \cite{Svetovoy2013}, where $\tau$ is the relaxation time for the charging-discharging process. For each pulse the function $I(t)$ is not concave that shows presence of the oxidation-reduction component. Due to this complication we roughly estimate the Faraday current using different procedure.

First, from the time-resolved data we determine the consumed power $P(U)$ averaged over 10 periods at a fixed amplitude of the voltage pulses $U$. From this power we extract the current $I_P=P/U$ shown in Fig.$\:$\ref{fig:current}(b). It can be fitted by the following nonlinear curve $I_P=64.2(U/U_0)^2+10.5(U/U_0)\:$mA, where $U_0=11\:$V is the maximal amplitude. The Faraday current gives the main contribution to $I_P$. We define it as $I_F=kI_P$, where the coefficient $k$ is estimated roughly as $k=0.67$ based on the observation that both the charging-discharging and the oxidation-reduction processes are faster than the pulse duration. Thus, at $U=U_0$ the Faraday current is estimated as $I_F=50 mA$. It is close to the value $I_F=60\:$mA reported for a similar but different sample \cite{Postnikov2017}. The Faraday current density is estimated as $j_F=70\:$A/cm$^2$ that is on the same level as for all alternating polarity experiments \cite{Svetovoy2016}. It is worth to note that slightly nonlinear behavior $I_P(U)$ is not related to the effect of heating of the electrolyte because a direct measurement of the temperature rise is smaller than $10^{\circ}\:$C (see below). This systematically observed effect for Ti electrodes (but not for Pt, Cu, or W) is apparently related to the long-time decrease of thickness of the oxide layer on Ti surface during the process.

Although no visible bubbles are produced by the alternating polarity process, significant Faraday current means that the gas is generated. This gas has to change the optical density of the medium nearby the electrodes. Using our schlieren setup we visualize how the optical density is distributed in space and time.

\subsection{Optical density}
\begin{figure}[tbhp]
\centering
\includegraphics[width=1.0\linewidth]{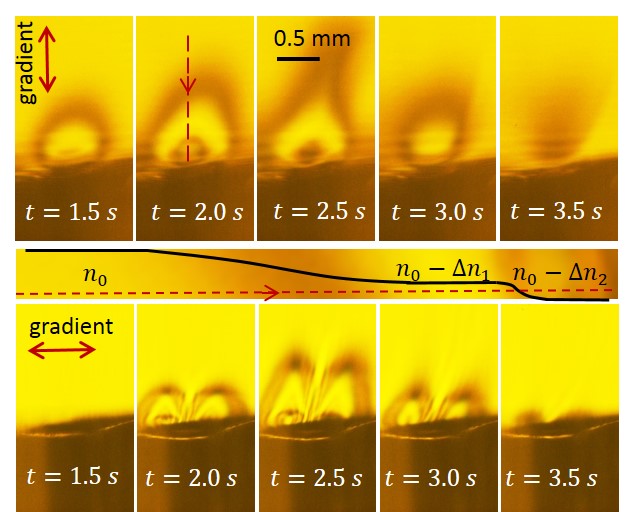}%Shade_50.jpg
\caption{Schlieren contrast produced by the alternating polarity process. The upper row (frames from Video S1) shows the vertical gradient of the optical density. The stripe below the images explains schematically formation of the contrast along the red dashed line in the image at $t=2.0\:$s (see the text). The lower row of images presents the gradient in the horizontal direction.}
\label{fig:shade_50}
\end{figure}

Square voltage pulses at frequency $f=200\:$kHz modulated by a triangle waveform with a period of $5\:$s are applied to the cell. The maximum voltage amplitude is $U_{0}=11\:$V. The schlieren contrast is shown in Fig.$\:$\ref{fig:shade_50} (see also Video S1). The upper row of images demonstrates successive changes of the gradient of optical density in the vertical direction. The lower row shows the changes in the horizontal direction.

Similar to the classical schlieren method our setup is sensitive to the gradient of the refractive index in the medium in the direction normal to the knife edge. The vertical gradient is resolved at $U>4.5\:$V. The region nearby the electrodes becomes darker as it should be for a liquid enriched with gas. When the voltage amplitude increases further the area of lower refraction grows and the central part becomes bright again as one can see in the frame at $t=1.5\:$s. In its central part the refraction index is smaller than in the background liquid  but homogeneous. Since there is no gradient we see this region on the same level of brightness as the background. The region of lower refraction forms a cloud of optically less dense liquid covering the electrodes.

In the frames corresponding to $t=2.0\:$s and $2.5\:$s a dark spot is visible in the center of the cloud. It originates from nonhomogeneous current density distribution, which is the largest at the edge of the central electrode, is somewhat smaller in the center, and is the smallest at the peripheral electrode. Because of this distribution the optical density above the central electrode is even smaller than in the main part of the cloud. This is clear from the fact that we see it dark again. If this region would be optically more dense, we would see a bright spot, not dark. With increasing voltage the area of smaller optical density increases but at the maximum voltage $U=11\:$V at $t=2.5\:$s one can see that the cloud develops a neck. This is an interesting phenomenon that will be discussed specially. One can notice that the neck is not exactly vertical. It is explained by uncontrolled convection that has probably the thermal origin. When the voltage decreases the system gets via the same states in the opposite direction. The contrast formation is explained schematically in stripe panel below the upper row of images in Fig.$\:$\ref{fig:shade_50}.

The lower row of images demonstrates approximate reflection symmetry of the electrodes. A ray passing through the center of the structure does not deviate but the ray passing left from the center deviates to the left from its original direction and the one passing right from the center deviates to the right as one can see in Fig.$\:$\ref{fig:scheme}(b). The images at $t=2.0\:$s and $t=2.5\:$s have a darker spot in the center of the left and right halves related to a higher concentration of gas near the edge of the central electrode.

From Fig.$\:$\ref{fig:shade_50} we can conclude that the application of the alternating polarity voltage pulses produces gas above the electrodes that does not form bubbles strongly scattering the visible light but reduces the refractive index of the liquid. This gas is distributed in the form of a dome covering the electrodes. The gas concentration within the dome is the same nearly everywhere but it is higher and distributed  inhomogeneously near the central electrode. At larger  voltage amplitude the dome develops a neck. Now we are going to discuss the origin of this neck.

\begin{figure}[tbhp]
\centering
\includegraphics[width=.99\linewidth]{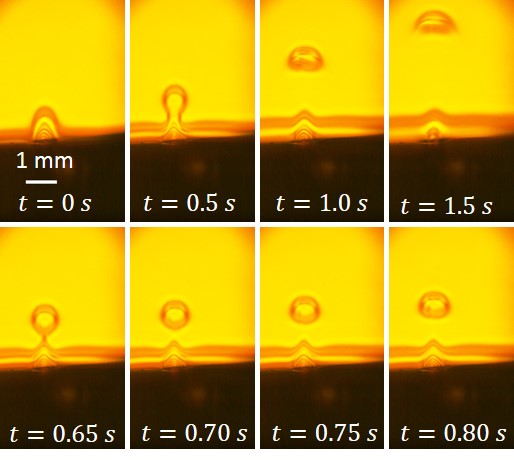}%Archimed.jpg
\caption{Schlieren contrast (frames from Video S2) for a larger field of view in comparison with Fig.$\:$\ref{fig:shade_50}. Upper row shows successive images separated by a time of $0.5\:$s. In the lower row evolution of the gas enriched "drop" is shown on the shorter time scale. The moment $t=0$ corresponds to the maximum voltage.}
\label{fig:archimed}
\end{figure}

To see the area above the neck we use an objective with $100\:$mm focal length. It enlarges the field of view. The same pulses at $200\:$kHz modulated by the triangle waveform are used. Figure \ref{fig:archimed} shows a few frames with schlieren contrast from Video S2. As the initial moment $t=0$ we take the state with the maximum voltage amplitude. The upper row shows a series of images taken with a time step of $0.5\:$s. We can see that initially the cloud with a lower refraction index is attached to the substrate as in Fig.$\:$\ref{fig:shade_50}. Then a volume of spherical shape is separated from the parent cloud but still is tied to it by the neck. This separated volume enriched with gas resembles a "drop" of liquid with a smaller mass density than the surrounding liquid. In contrast with a normal drop it has no sharp boundary. In the next image at $t=1.0\:$s the "drop" is completely detached and rising up to the free surface of the liquid. In the last image of the upper row one can see that the "drop" changes its shape during rising as would do a normal drop of lighter liquid or a bubble. When the "drop" reaches the free surface the gas does not go out to the atmosphere but spread along the free surface as one can see in Fig.$\:$S1 (see Supplemental).

In the lower row four successive images (time step is $50\:$ms) are shown. They demonstrate how the "drop" rises up preserving its shape on this short time scale. From the last three images at $t=0.70,\ 0.75$ and $0.80\:$s one can estimate the rising velocity as $v=2.8\:$mm/s and the "drop" radius as $R=315\:\mu$m. Physically we have got here a nearly spherical volume enriched with gas. The mass density in this volume is smaller than the density in the background liquid. Therefore, the "drop" has to be subjected to buoyancy that explains its rise. When the sample is mounted parallel to the gravity direction, one can see that we really deal with the buoyancy force (see Fig. S2 in Supplemental).

\subsection{Thermal effect}
Alternatively the observed contrast could be generated by a nonhomogeneous temperature distribution around the electrodes.
The current flowing through the electrolyte produces Joule heat that can change the refractive index of the liquid. Direct measurement of the temperature rise nearby the electrodes with the thermocouple gives the maximum value $\Delta T=10\pm 1\:^{\circ}\:$C just above the central electrode. No temperature change is found at a distance of $3\:$mm from the center of the electrodes in any direction.

\begin{figure}[tbhp]
\centering
\includegraphics[width=.99\linewidth]{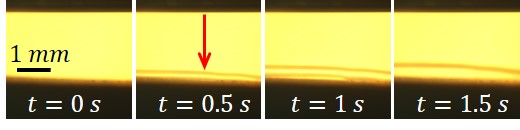}%Heat.jpg
\caption{The thermal contrast (frames from Video S3). The arrow in the second  image indicates position of the heater center. Triangle voltage with a period of $2\:$s is applied to the heater.}
\label{fig:heat}
\end{figure}

To see how the temperature distribution influences the schlieren contrast we observe the contrast produced by the heater (see Fig.$\:$\ref{fig:electrodes}(c)) in distilled water. Applying a triangle voltage from 0 to $7.9\:$V with a period of $2\:$s we observe the contrast presented in Video S3. A few frames from this video are shown in Fig.$\:$\ref{fig:heat}. The temperature rise of the heater loop is estimated as $\Delta T=9.5^{\circ}\:$C at maximum voltage. The maximum power dissipated by the loop is $P_l=0.9\:$W that is close to the power $P_{el}=0.8\:$W dissipated in the electrochemical process. The total maximum power of the heater $P_{tot}=4.4\:$W is larger due to heat produced by the contact lines. These lines also give contribution to the contrast although the extra power is dissipated far away from the loop.

One can see a completely different character of the thermal contrast. The region with lower optical density is broader in the horizontal direction but it is not going far from the substrate in the vertical direction. The latter is related to a high heat conductivity of the silicon substrate. The observed movement of the contrast is explained by the convection along the substrate. It has to be noted that there is no a preferable direction of the convection that is randomly defined from run to run. Now we can relate a part of the contrast observed during the electrolysis with the thermal effect. In Video S2 (see also Fig.$\:$\ref{fig:archimed}) one can see that with the voltage increase and decrease appears and disappears the contrast concentrated in a layer above the substrate.

\subsection{Size of the bubbles}
We already established that the optical contrast appears due to liquid enriched with gas. This gas has to form bubbles, whose sizes can be determined using the light scattering. The bubbles of a size comparable with the wavelength of visible light can be easily identified due to strong scattering of light. However, we do not observe these bubbles independently on the generation or observation time. Therefore, one would expect that the gas that manifests itself as the optical contrast has to form NBs. The latter weakly scatter the light but their size can be determined using the DLS method.

\begin{figure}[tbhp]
\centering
\includegraphics[width=.95\linewidth]{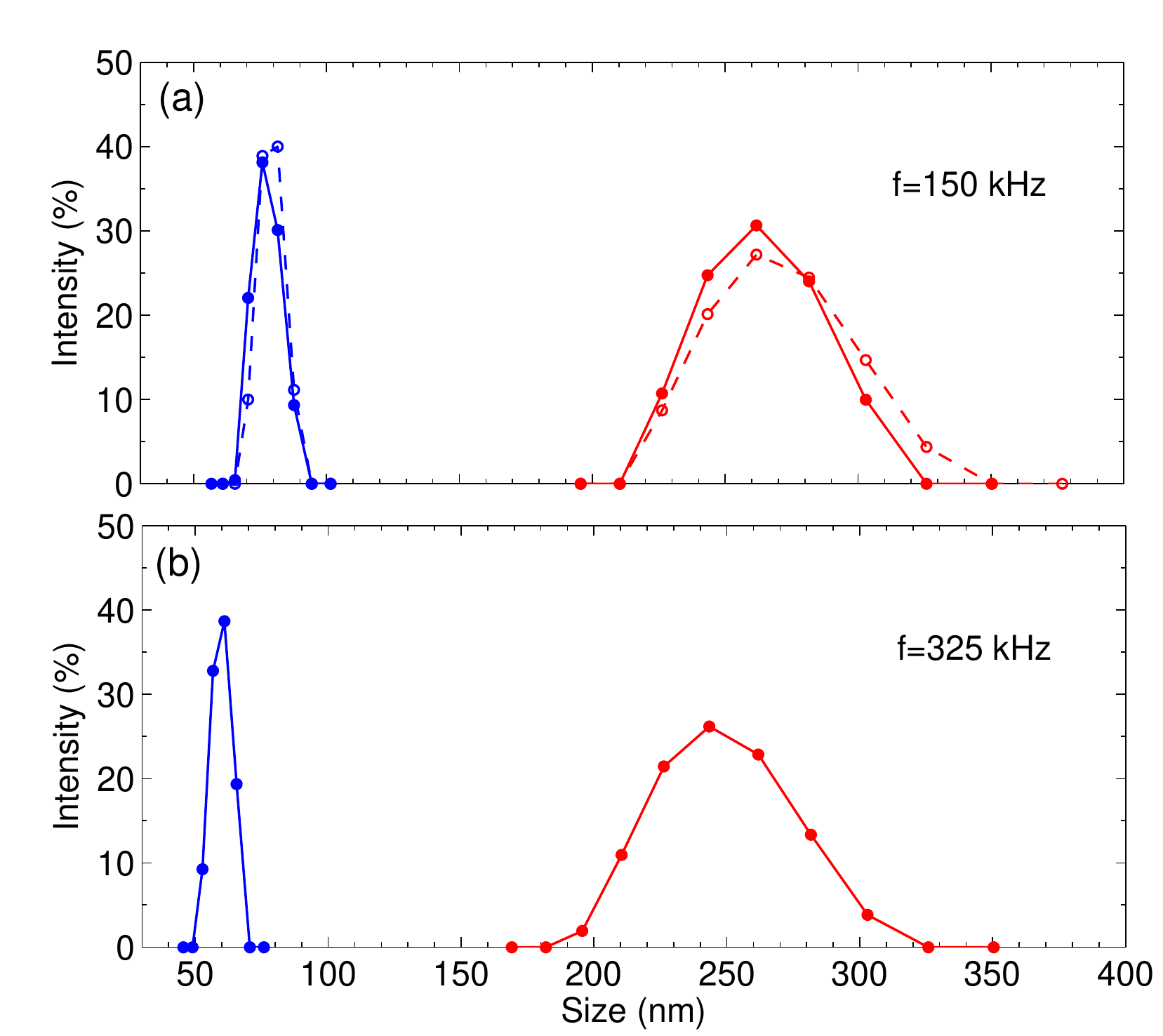}%DLS_size.pdf
\caption{Size distribution of NBs. The blue curves (smaller sizes) correspond to the data collected when the electrical pulses are switched on. The red curves show the size distribution $3\:$min after switching off the pulses. Panels (a) and (b) show the data collected for different frequencies of the pulses.  }
\label{fig:DLS_size}
\end{figure}

Investigation of light scattering with Zetasizer nano ZS analyzer reveals that the NBs are indeed generated by the alternating polarity process. After careful filtering of the solution the analyzer does not identify any nanoparticles in the solution. When the electrical pulses are switched on we observe bubbles with the diameter in the range of $50-90\:$nm. After switching off the pulses the signal becomes weaker and disappears in $15\:$min. What is interesting is that the bubble sizes that are observed after the process lie in the range of $200-300\:$nm. They are larger and their distribution is broader than those during the process. The change in the sizes happens faster than the time needed to collect the signal ($\sim 1\:$min) so that we are unable to see the transition.

Some results are presented in Fig.$\:$\ref{fig:DLS_size}. In panel (a) two different runs for the same voltage amplitude $U=11\:$V and frequency $f=150\:$kHz are shown. These two runs demonstrate reproducibility of the size distribution. The peak at smaller bubble diameter is found during the process and the peak at larger diameter corresponds to the measurement that is done $3\:$min after the process. Panel (b) shows the signal for $U=11.5\:$V and $f=325\:$kHz. In this case the bubbles are slightly smaller but have very similar widths of the distributions. With the available data we cannot make a definite conclusion that the bubble size depends on the frequency.

\subsection{Discussion}
Our results demonstrate that the alternating polarity electrochemical process generates NBs around the electrodes. No microbubbles strongly scattering the visible light are formed in contrast with normal DC electrolysis. Although we do not identify the gases responsible for the change of the optical density, a significant Faraday current shows that these gases are hydrogen and oxygen. According to the Faraday law the total number of gas molecules produced during a 5 second cycle is
\begin{equation}\label{N_Faraday}
  N=\frac{3}{4|e|}\int_0^{t_0}I_F(t)dt\approx 4\times 10^{17},
\end{equation}
where $t_0=5\:$s is the period of the triangle modulation and $e$ is the charge of electron. Even if all this gas will be homogeneously distributed in the electrolyte (total volume of about $1\:$ml), the liquid will be supersaturated with both gases already after one cycle.

In reality most of the gas is concentrated near the electrodes providing the contrast presented in Fig.$\:$\ref{fig:shade_50}. Although with our schlieren setup we cannot estimate the gas concentration in the cloud, one can do this with the optical distortion method. In this experiment, when the sample is observed from the top, we see optical distortion of the electrodes visible by naked eye. This effect was described earlier \cite{Postnikov2017} and the gas concentration responsible for the distortion was estimated. In this experiment the number of gas molecules produced by the process is $2-3$ times smaller and we expect the gas concentration in the cloud as high as $n_g\sim 1\times 10^{26}\:$m$^{-3}$.

The lump of NBs ("drop") that is separated from the parent cloud rises up because it has a smaller mass density than the surrounding liquid. We can estimate a volume fraction of gas $f_g$ in the "drop" using as input parameters the radius of the "drop" $R$ and its velocity $v$. Equating the buoyancy force and the Stokes' drag acting on the "drop" one finds for the gas fraction
\begin{equation}\label{g_frac}
  f_g=\frac{9\eta v}{2(\rho_l-\rho_g)gR^2},
\end{equation}
where $\rho_l$ and $\rho_g$ are the densities of liquid and gas, respectively, $g$ is the free-fall acceleration, and $\eta$ is the viscosity of the solution. From this expression we find $f_g\approx0.013$. Since the lump exists separately from the parent cloud, it has to be composed of larger NBs with a size of $250\:$nm. Calculating the total number of gas molecules $N$ in the "drop" with the help of the ideal gas law one finds for the gas concentration $n_g\approx 4\times 10^{24}\:$m$^{-3}$. It is much smaller than that in the parent cloud since the "drop" is not supported by the gas produced by the electrodes. Additionally, the "drop" is composed from larger NBs that have smaller Laplace pressure.

Observation of different sizes of NBs during and after the electrochemical process can be explained by the Ostwald ripening. The cloud is composed of densely packed hydrogen and oxygen NBs. The concentration of NBs is $n_{NB}=n_g/N_r$, where $N_r$ is the number of gas molecules in one NB. The latter can be estimated using the ideal gas law. Thus, for $n_g=1\times 10^{26}\:$m$^{-3}$ and for the average bubble radius $r=40\:$nm one finds $n_{NB}=4\times 10^{20}\:$m$^{-3}$. For this concentration of NBs the average distance between the bubble centers is estimated as $140\:$nm so that the bubbles are separated only by the distance of $60\:$nm. At such a small separation the bubbles exchange fast by the content via the gas diffusion in liquid. If two neighboring bubbles have the same gas, one bubble disappears while the other one grows. If the gases in the neighboring bubbles are different, then one bubble disappears but the other one becomes smaller. This is due to combustion reaction between hydrogen and oxygen that happens in NBs spontaneously at room temperature \cite{Svetovoy2011}. A steady state in bubble distribution within the cloud is supported by the gas produced by the electrodes.

When the electrochemical process is switched off the size distribution is shifted to a larger values and the total gas concentration is reduced. We observe the NBs $15\:$min after switching off the process. This time is much longer than the time $\sim 1\:$ms needed for diffusive dissolution of a $250\:$nm bubble \cite{Epstein1950}. On the other hand, it is shorter than the lifetime of bulk NBs produced mechanically (days) \cite{Ohgaki2010,Ushikubo2010} or by normal electrolysis (hours) \cite{Kikuchi2001}. The shorter lifetime of NBs in our experiment is explained not by different physical properties of the bubbles but rather by the possibility of the gases to interact: occasionally two NBs with different gases can meet and disappear in the reaction.

The lump of NBs separated from the parent cloud is not supported by the gas from the electrodes. For this reason it has to contain larger NBs with an average size of $250\:$nm. The concentration of NBs in the "drop" is estimated as $n_{NB}=2\times 10^{18}\:$m$^{-3}$ and the average distance between them equal to $0.8\:\mu$m is much larger than that in the parent cloud.

\section{Conclusions}
We generated a dense cloud of H$_2$ and O$_2$ nanobubbles that was controlled in space and time.  It has been done using the electrochemical decomposition of water with short voltage pulses of alternating polarity. The process produces only small NBs which do not scatter strongly the visible light. Detailed structure of the cloud was investigated using collective effects produced by the NBs. Significant fraction of the gas in the liquid was visualized using lens and grid schlieren setup. The NBs formed a dome-shaped cloud covering the concentric electrodes. Inside of the dome the NBs are distributed inhomogeneously in correspondence with the current density distribution.

We observed the separation of a lump of NBs from the parent cloud. The lump takes a spherical shape and rises up to the free surface changing its shape on the way. One NB has a negligible buoyancy but the lump was moved as a whole by the gravity since it has smaller mass density than the surrounding liquid. From this movement we estimated the density of NBs in the lump as $2\times 10^{18}\:$m$^{-3}$, which is considerably smaller than that in the parent cloud.

Using the dynamic light scattering we determined the size and lifetime of NBs in the cloud. It is the first direct measurement of the sizes of NBs in the alternating polarity electrochemical process. While the electrical pulses were switched on, the average size of the bubbles was found to be $60-80\:$nm, but when the pulses were switched off the distribution became more broad and shifted to larger values around $250\:$nm.  These bubbles can be observed 15 minutes after the process has been terminated. The change in the distribution and the concentration of NBs during and after the electrochemical process was related to the spontaneous combustion reaction between hydrogen and oxygen in NBs.

\section*{Acknowledgments}
This work is supported by the Russian Science Foundation, grant No.15-19-20003.

%%%REFERENCES%%%
%\bibliography{Shade_lib}


\end{document}